\documentstyle[preprint, aps]{revtex}

\draft

\begin{document}


\title{Absence of ``Floating'' Delocalized-States in a Two-Dimensional Hole Gas}

\author{S.~C.~Dultz and H.~W.~Jiang} 
\address{Department of Physics and Astronomy, University of California
  at Los Angeles, Los~Angeles,~CA~90095}

\author{W.~J.~Schaff}
\address{Department of Electrical Engineering, Cornell University,
  Ithaca,~NY~14853} 

\date{\today}

\maketitle

\begin{abstract}
  By tracking the delocalized states of the two-dimensional hole gas in a p-type GaAs/AlGaAs heterostructure as a function of magnetic field, we mapped out a phase diagram in the density - magnetic field plane.  We found that the energy of the delocalized state from the lowest Landau level flattens out as the magnetic field tends toward zero. This finding is different from that for the two-dimensional electron system in an n-type GaAs/AlGaAs heterostructure where delocalized states diverge in energy as $B$ goes to zero indicating the presence of only localized states below the Fermi energy.  The possible connection of this finding to the recently observed metal-insulator transition at $B = 0$ in the two-dimensional hole gas systems is discussed.
\end{abstract}


\narrowtext

The transport properties of the two-dimensional hole gas (2DHG) in GaAs/AlGaAs heterostructures was one of the well-studied topics in the early 80s.  The focus was placed on the high magnetic field regime where the physics is predominantly determined by the quantum Hall effect \cite{Mendez}.  Very recently, there has been renewed interest in the 2DHG.  This interest is largely inspired by the experimental findings \cite{Hanein,Simmons} of a metal to insulator transition at zero magnetic field.  In these new experiments, a metal to insulator transition was found when the density of the 2DHG decreased below a critical value.  These results are very similar to that observed in the Si MOSFET system \cite{Kravchenko} and are contradictory to the well-known scaling theory \cite{Abrahams} which predicted that there is no metallic state in two-dimension.  On the other hand, the two-dimensional electron gas (2DEG) in GaAs/AlGaAs heterostructures shows no such transition at $B = 0$, and its behavior can be described well by the scaling theory.\cite {Van Keuls}  This apparent difference is attributed to the strong Coulomb interactions in the 2DHG system \cite{Hanein}, and there are recent theoretical suggestions \cite{Dobrosavljevic} that various unusual ground states can be formed due to the strong Coulomb interaction.  There are indeed important physical difference between the hole system and the electron system.  First, due to the heavier effective mass of the holes, the dimensionless parameter r$_{s}$, which measures the ratio of the correlation energy to the Fermi energy, is much larger for the 2DHG.  Secondly, spin is more important in the hole gas system because of the breaking of spin degeneracy of the heavy hole band which is stronger in this valence band than in the conduction band.  The heavy hole band is split by the asymmetric well into the heavy-heavy hole (m$_{j}$ = 3/2) and the light-heavy hole (m$_{j}$ = -3/2) which contain holes of opposite spin.

In these new experiments \cite{Hanein,Simmons}, the temperature dependence of the resistivity  at $B = 0$ has been studied in detail to determine both the metallic and the insulating phases.  It was assumed that a negative temperature coefficient (d$\rho/dT<0$) was the signature for an insulating phase and that a positive temperature coefficient revealed a metallic phase.  While this method is simple and powerful in determining the insulating phase, a positive identification of the metallic phase can be difficult particularly when the temperature dependence is weak.  In the Cambridge group study \cite{Simmons}, in fact, the temperature dependence in the identified metallic phase is very weak and is found to be saturated at low temperatures.  More seriously, to identify a metallic phase by the positive temperature coefficient could be very misleading.  It is a well known fact that the resistivity drops continuously at low temperatures even for a high mobility 2DEG (i.e., a system with small r$_{s}$) due to the reduction of scattering at low temperatures.  \cite{Pfeiffer}  To determine whether the electronic wavefunction is localized in the thermodynamic limit (i.e., an infinitely large sample), very low temperatures are required for a system with a large correlation length.  Very often, the low-temperatures are not experimentally accessible.

There is an alternative way to determine whether there are delocalized states at $B = 0$ by tracking the energy position of the delocalized states in the magnetic fields.  It is known experimentally that the energy of the delocalized states are centered around the Landau levels in high magnetic field.  By following the evolution of their position as $B\rightarrow0$, one can determine whether there are delocalized states at $B = 0$ in the thermodynamic limit.   This method appears to be effective.  For example, it was found that for a 2DEG in a GaAs/AlGaAs heterostructure the position of the delocalized states did not remain at the Landau level centers in the limit of low magnetic field.  Instead, they rise up in energy, piercing the fermi surface at some finite magnetic field and tend to infinity in the limit of very low magnetic field.\cite{Glozman}  This finding is consistent with the understanding that there are no 2D metallic states in a thermodynamic limit.  In the present study, we would like to employ the same experimental method to determine whether there are delocalized states below $E_{F}$ of the 2DHG.

The samples used in this experiment were cut from a p-type GaAs/AlGaAs wafer fabricated by molecular beam epitaxy.  A $150~$\AA~Al$_{.45}$Ga$_{.55}$As undoped spacer was used, followed by a $500~$\AA~Be-doped ($1~\times~10^{18}\text{~cm}^{-3}$) Al$_{.45}$Ga$_{.55}$As layer.  The mobilities are 35,000 $cm^{2}/V-sec$ for sample A and 50,000 $cm^{2}/V-sec$ for sample B with hole densities of $5.6~\times~10^{11}\text{~cm}^{-2}$ and $3.4~\times~10^{11}\text{~cm}^{-2}$ respectively.  A Hall-bar pattern was etched into each sample using standard photolithography techniques with dimension 500 $\mu$m$~\times~$1500 $\mu$m for sample A and 150 $\mu$m x 450 $\mu$m for sample B.  NiCr gates were evaporated on the top so that applying a voltage with respect to the drain could vary the carrier density of the 2DHG.  In:Zn diffusion was used for contacts with a 20\% concentration of Zn by weight.  The samples were thermally connected to the mixing chamber of our dilution fridge.  Standard low-frequency lock-in techniques were used for the magnetotransport measurements.

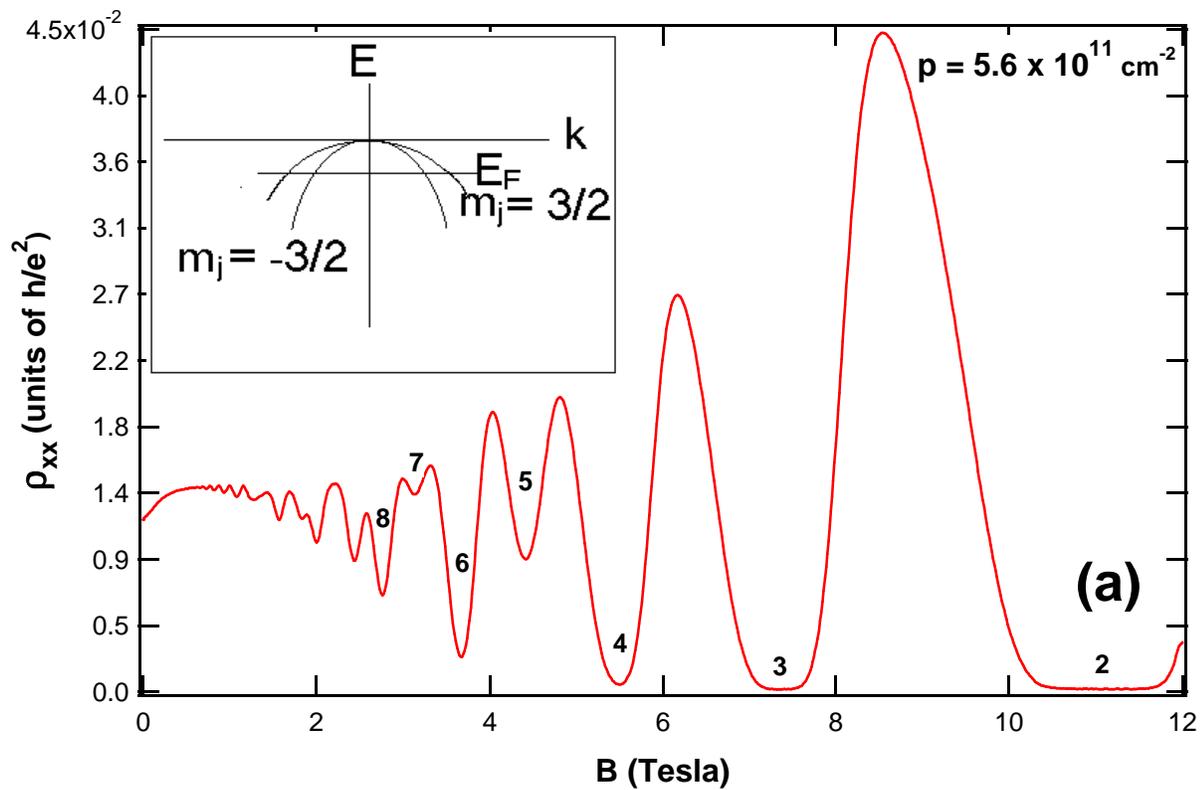
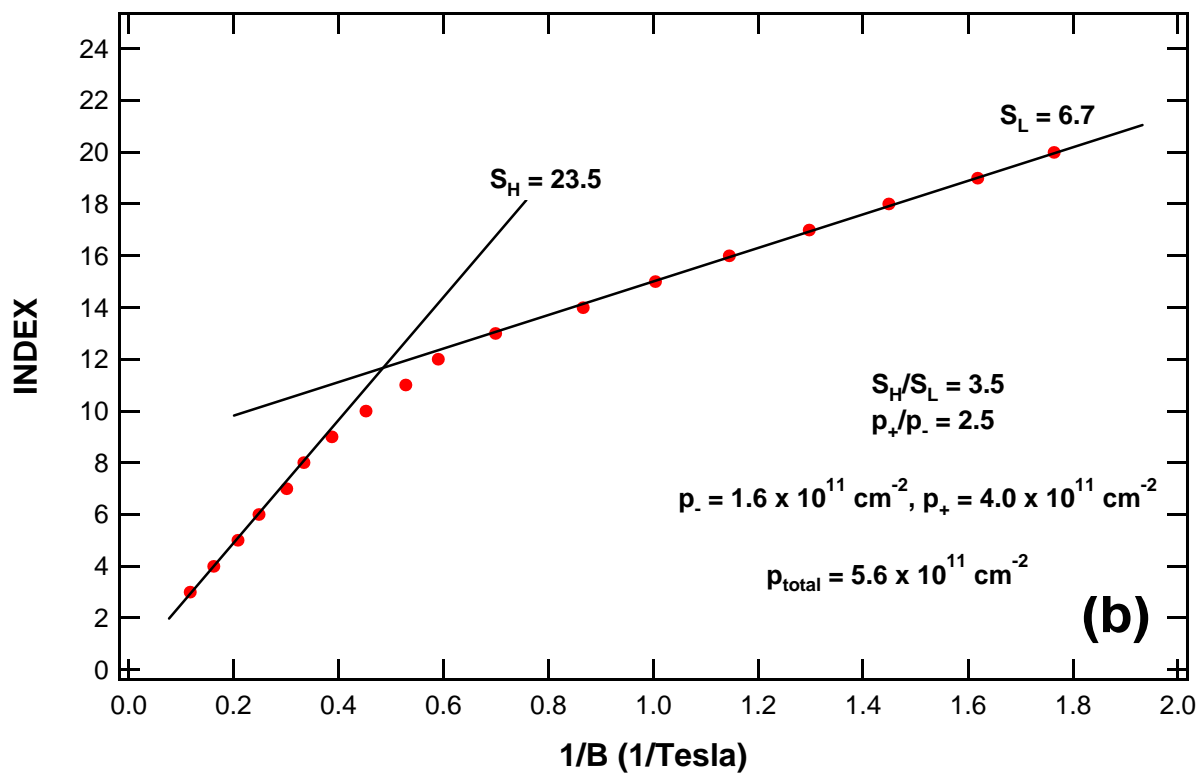
\begin{figure}
  \caption{(a) A typical trace of the magnetoresistance for the 2DHG.  Inset shows the
  situation when both spin-split valence bands ($m_j$ = 3/2 and $m_j$ = -3/2) are filled.
  (b) the index of the peaks is plotted against the position (1/B) showing two different
  slopes which represent the presence of the spin subbands.  The densities of the two bands
  are deduced from these slopes.}  
\label{QH}
\end{figure}

Fig. \ref{QH}a shows a typical trace of $\rho_{xx}~vs.~B$ in the high density regime for the 2DHG.  As indicated by Fig. \ref{QH}b (the peak index vs. $1/B$), there are two sets of oscillations.  These two sets of oscillations are due to the fact that both the $m_j$ = 3/2 and $m_j$ = -3/2 bands are filled as schematically shown in the inset of Fig. 1a.  From such a plot, the densities of the $m_j$ = 3/2 band ($p_+$) and the $m_j$ = -3/2 band ($p_-$) can be deduced \cite{Stormer}.  The energy of the spin splitting at $k_F$ and at zero field is estimated to be 1.32 meV at that particular density.  More importantly, the Fermi energy for this study can also be obtained.  It is worth noting here that the Fermi energy in the parabolic band approximation can be expressed as $\frac{\pi\hbar^{2}\text{p}}{m*}$ where ${m*} = 0.49m$ and $p_{total}$ = ($p_+$ + $p_-$).  This energy should be twice that calculated from $p_+$ and $m_+$ ($E_F$ = $\pi\hbar^{2} p_+/m_+$), or from $p_-$ and $m_-$ ($E_F$ = $\pi\hbar^{2}p_-/m_-$).  We don't write explicit values for $m_+$ and $m_-$ because these masses are also density dependent but ${m*} = 0.49m$ is density independent.
 	
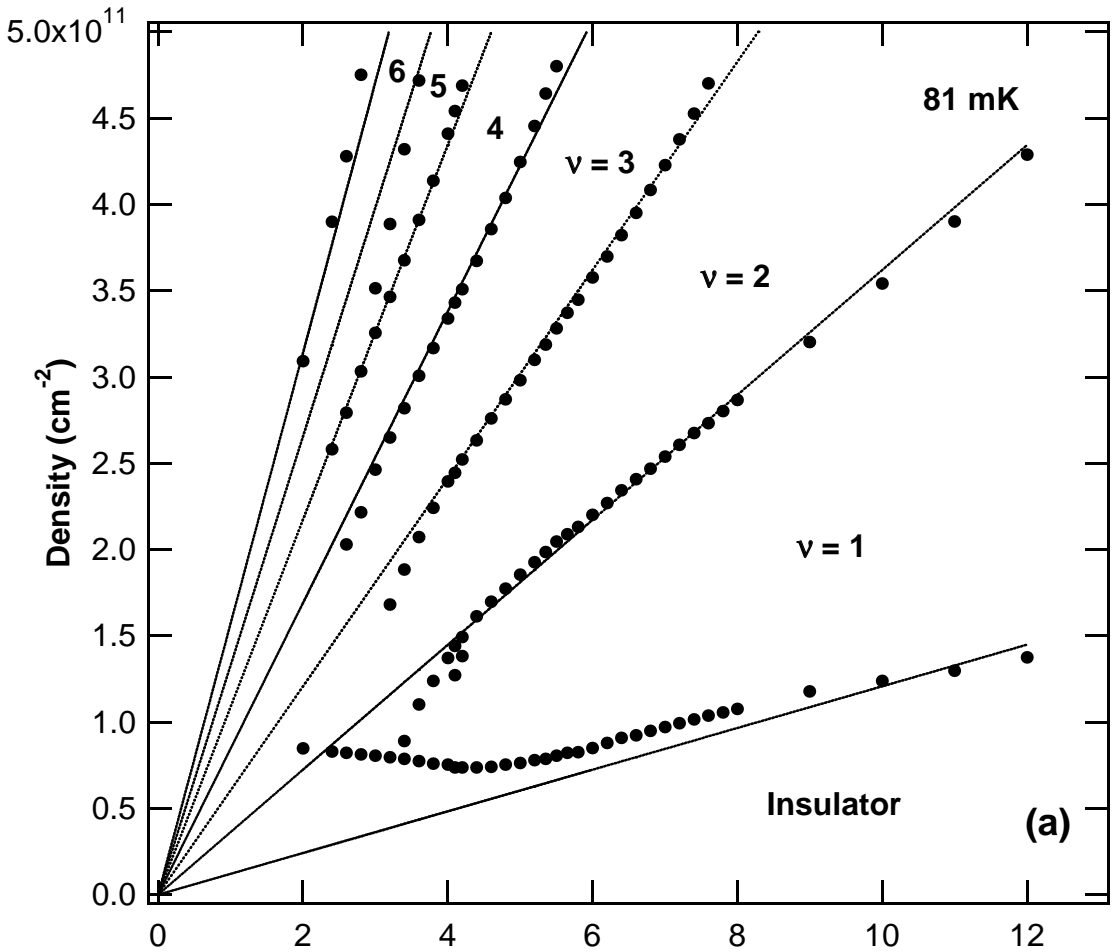
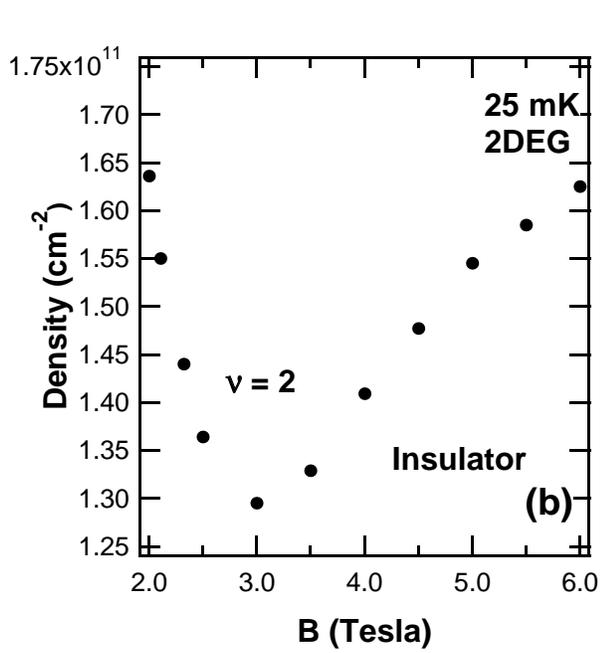
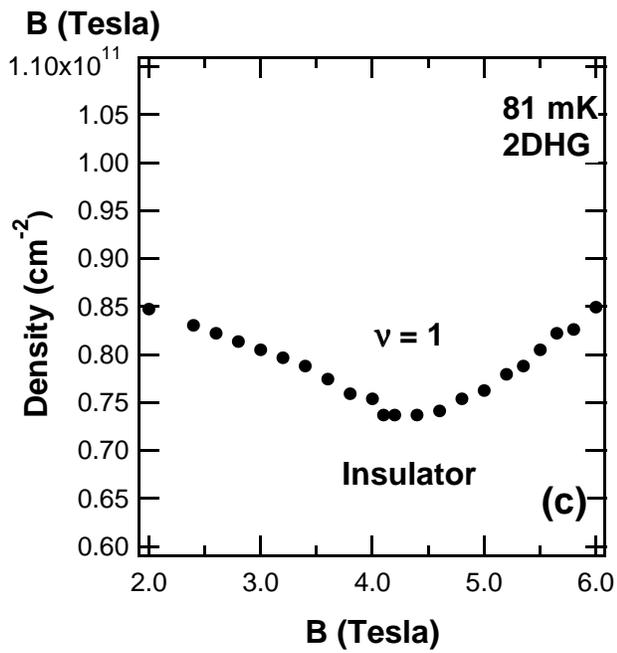
\begin{figure} 
  \caption{(a) A map of the delocalized states in the $n-B$ plane for the 2DHG system.
  (b) and (c) show the sharp contrast between the 2DEG and the 2DHG for the delocalized
  states of the lowest Landau level.  Note the energy of the delocalized states for the 
  2DEG goes up continuously as $B\rightarrow0$ (floating) while that for the 2DHG is flattened
  out for small $B$.}
\label{Phase}
\end{figure}

Fig. \ref{Phase}a is the experimental phase diagram of the insulator - quantum Hall liquid transition for Sample A (higher density sample).  The data points represent the position of the phase boundaries (the delocalized states) in the $n-B$ plane.  The diagram is obtained from traces of $\sigma_{xx}$ as a function of applied gate voltage taken at a fixed magnetic field.  Each data point represents a distinct peak in $\sigma_{xx}$ or correspondingly to a temperature independent crossing point in $\rho_{xx}$ (see Fig. \ref{IMI} for example).  A single trace therefore marks a vertical slice of this phase diagram where only a few peaks might be present as we scan the full range of gate voltage until the sample eventually becomes insulating at the lowest densities.  As the magnetic field is lowered, the peaks are becoming increasingly broad as limited by the base temperature of our fridge of 60 mK.  As mentioned earlier, the finite sample size (i.e., small inelastic length) due to finite temperature can disrupt the determination of the metallic phase.  We therefore stop the data processing at about $B = 2$ T where the error bars of the peak position are too large to provide useful information on the absolute position.  We would like to mention, however, that the error bars are on the order of the dot size in figure \ref{Phase}a at $B = 2$ T but get much smaller as the magnetic field in increased.  The corresponding total density $p_{total}$ is found from Shubnikov De Haas oscillations and the Landau level centers ($p = (n + \frac{1}{2}$)$\frac{eB}{hc}$) are plotted (dashed lines) as a check for the correct densities.

In high fields, delocalized states follow the Landau level centers as should be expected both for the 2DEG and for the 2DHG.  However, this phase diagram at the low-field regime is qualitatively different than the one obtained from the 2DEG \cite{Glozman}.  The most striking difference is the flattening out (i.e., delocalized state terminated at a finite density) of the lowest delocalized state in lower and lower field.  In the 2DEG data as shown in Fig. \ref{Phase}b, the lowest delocalized state is a minimum at around 3 T and begins to come up sharply, apparently diverging before disappearing in the lowest fields.  This continuous rising in chemical potential as $B$ tends to zero represents the ``floating'' \cite{Khmelnitskii} of the delocalized states above the Fermi level.   Naturally, a metal to insulator transition would be impossible.  There would only be localized states below the Fermi level no matter what the density was in a thermodynamic limit.  In contrast, for the present experiment, the minimum occurs at around 4.2 T but then rises to a shallow, apparently linear slope rather than diverge (shown in a magnified view in Fig. \ref{Phase}c).  Notice that both (b) and (c) are plotted over the same range of density to give a fair comparison.  To compare the results more quantitatively, we have taken derivatives of these data.  We found that the derivative of the data for the 2DEG gets increasingly negative throughout the trace.  However, for the 2DHG, the derivative turns around and never gets steeper than $-7~\times~10^{9}\text{~cm}^{-2}$ $T^{-1}$ as $B\rightarrow0$ which is about an order of magnitude smaller at 2 T than that from the 2DEG data.  

We believe this flattening out implies that it is possible for delocalized states to exist at $B = 0$ for a sufficiently high Fermi energy.  One can vary the gate voltage of the sample, thereby changing the density of the hole gas and be able to tune the Fermi level above and below the lowest delocalized state energy.  This could therefore conceivably produce a metal to insulator transition.  In fact, it was observed by Shashkin et al. \cite{Shashkin} in a Si MOSFET, that the position of the delocalized states (determined by a resistivity cut-off method) levels off towards a finite density as $B\rightarrow0$.  And it is now commonly accepted that there is indeed a metal-insulator transition in the Si MOSFET.  The critical density extrapolated from the figure would be $p_c$ = $1.0\pm0.1 \times 10^{11} \text{cm}^{-2}$ (corresponding to $r_s$ = 13.7).  There is another apparent difference between the two systems.  For the 2DHG, the minimum of the delocalized states occurs at the boundary between the $\nu$ = 1 quantum Hall liquid and insulating phase.  For the 2DEG, the minimum occurs at the boundary between the $\nu$ = 2 spin-degenerate quantum Hall liquid and the insulating phase.  We believe this difference is a result of the strong spin splitting in the 2DHG system.  It is possible that the energy of the splitting is comparable to the cyclotron energy even at a few Tesla.  

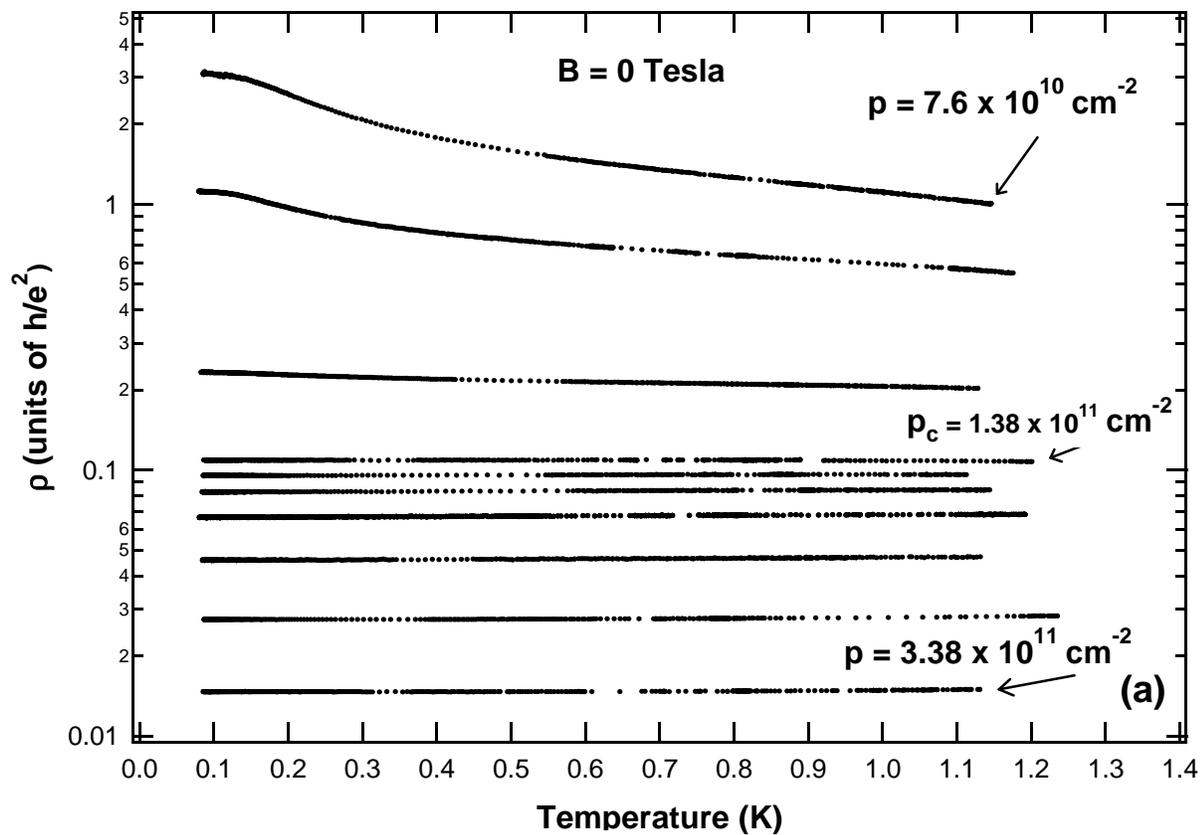
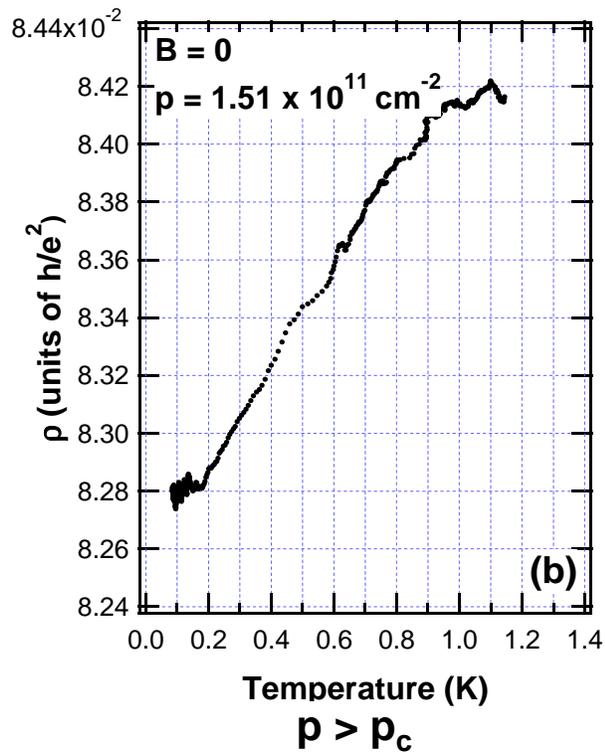
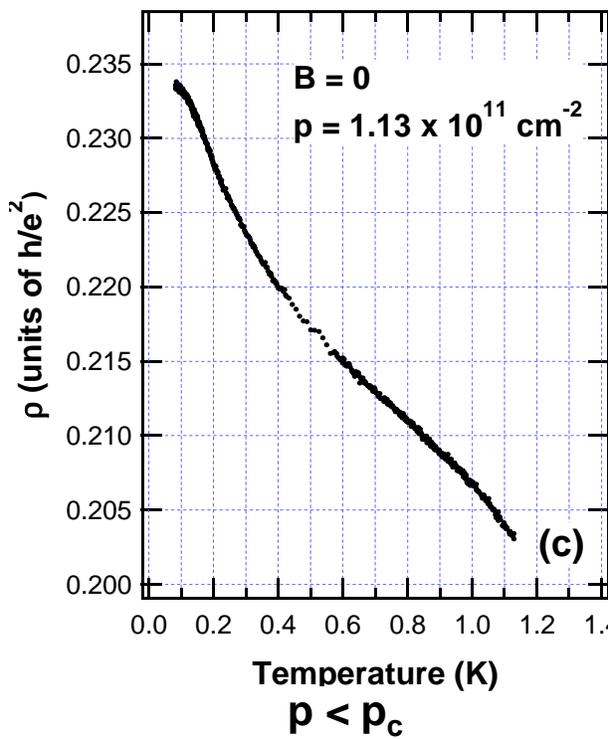
\begin{figure}
  \caption{(a) The temperature dependence of the resistivity for a sequence of hole
  densities at $B = 0$.  Densities from the top to bottom trace and in units of $10^{11}   {cm}^{-2}$ are as follows:  0.76, 0.88, 1.13, 1.38, 1.44, 1.51, 1.63, 1.88, 2.38, 3.38.  (b)   and (c) are magnified views of the $p = 1.51\times10^{11} {cm}^{-2}$ and $p = 1.13\times10^{11}   {cm}^{-2}$ traces respectively.  Note the difference in the sign of the temperature   coefficient.}  \label{Rho}
\end{figure}

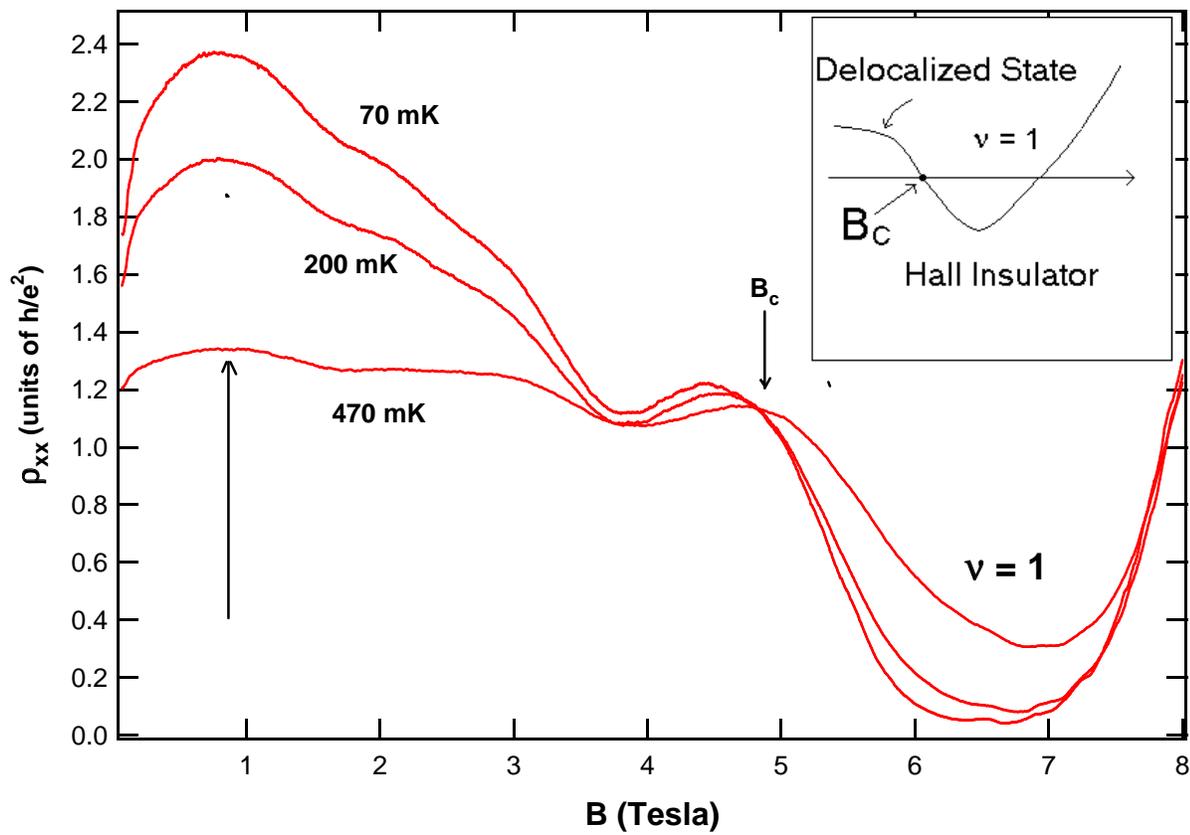
\begin{figure} 
  \caption{$\rho_{xx}~vs.~B$ for three temperatures.  The crossing point of the three
  curves corresponds to the critical magnetic field when the Fermi energy crosses the 
  delocalized states as shown schematically in the inset.  Also note the peak in $\rho_{xx}$
  at $B = 0.8$ T (discussed in the text).}
  \label{IMI}
\end{figure}

Next we would like to discuss the possible connection of our observation with the results yielded from the temperature dependence of the resistivity studies \cite{Hanein,Simmons}.  Figure \ref{Rho}a shows resistivity vs. temperature traces at $B = 0$ for our lower density sample.  The temperature dependence is very weak in the high hole densities (definitely much weaker than that seen in the Si MOSFET and similar to that of the 2DHG \cite{Simmons}).  A zoomed view of Fig. \ref{Rho}a, however, clearly shows the positive temperature coefficient which was used as evidence for a metallic phase \cite{Hanein,Simmons}.  A switch of the positive to negative coefficient occurs at a density of p = $1.38 \times 10^{11}\text {cm}^{-2}$ (corresponding to $r_s$ = 11.8).  Although this density is not exactly the same as that obtained by the delocalized states mapping study described above, there are very close considering the wide range of the density covered for the study.  We believe that the result determined by the temperature coefficient study should be valid in determining the crossover density in the thermodynamic limit for this system.

In considering the unusually large spin-split due to the lack of inversion symmetry, there is a natural question whether the suppression of the floating is a result of the spin-orbital scattering.  It has been proposed recently that the spin-orbital interaction can change the normal orthogonal symmetry of the system and therefore induce a metallic phase \cite{Pudalov}.  We would like to add a note here on the possible relevance of the spin-orbital scattering.  We have found that the magneto-resistance data at very low-fields consistently show a positive magneto-resistance which reaches a peak always at around $B = 1$ T.  Fig. \ref{IMI} shows such a trace in the insulating phase.  A pronounced peak can be seen at $B = 0.8$ T (indicated by the arrow), followed by a sharp drop (negative magnetoresistance).  It is known that a low-field positive magnetoresistance can be caused by the spin-orbital scattering in both the diffusive regime \cite{Dresselhouse} and in the hopping conduction regime \cite{Shahbazyan}.  To determine quantitatively how much of the spin-orbital scattering suppresses the floating,  additional experiments are required.

In summary, we have presented an alternative way to determine the existence of a metal to insulator transition in the 2DHG in the GaAs/AlGaAs heterostructure by tracking the delocalized states in the $n-B$ plane.  Our experiments indicate that the energy of the delocalized state from the lowest Landau level flattens out as $B$ tends to zero in sharp contrast to the 2DEG GaAs/AlGaAs system.  This finding suggests that there exist delocalized states above a finite density in a thermodynamic limit at $B = 0$.
 
The authors would like to thank S. Chakravarty and S. Kivelson for helpful discussions.  This work is supported by NSF under grant \# DMR 9705439.

\end{document}